\documentclass[review]{elsarticle}

\usepackage{lineno,hyperref}
\modulolinenumbers[5]

\journal{Journal of \LaTeX\ Templates}
\usepackage[T1]{fontenc}
\usepackage{ae,aecompl}
\usepackage{color}

\usepackage{graphicx}	% Including figure files
\usepackage{amsmath}	% Advanced maths commands
\usepackage{amssymb}	% Extra maths symbols

\begin{document}

\begin{frontmatter}

\title{Reconstructing equation of state of dark energy with principal component analysis}
\tnotetext[mytitlenote]{Fully documented templates are available in the elsarticle package on \href{http://www.ctan.org/tex-archive/macros/latex/contrib/elsarticle}{CTAN}.}

\author[mymainaddress]{Zhi-E Liu}


\author[mysecondaryaddress]{Hao-Feng Qin}
\author[mymainaddress]{Jie Zhang}
%\author[mysecondaryaddress]{Tong-Jie Zhang\corref{mycorrespondingauthor}}
\author[mymainaddress,mysecondaryaddress,forzhang]{Tong-Jie Zhang\corref{mycorrespondingauthor}}
\cortext[mycorrespondingauthor]{Corresponding author}
\ead{tjzhang@bnu.edu.cn}
\author[foryu]{Hao-Ran Yu}

\address[mymainaddress]{College of Physics and Electronic Engineering, Qilu Normal University, Jinan, 250200, China}
\address[mysecondaryaddress]{Department of Astronomy, Beijing Normal University, Beijing, 100875, China}
\address[forzhang]{Institute for Astronomical Science, Dezhou University, Dezhou 253023, China}
\address[foryu]{School of Physics and Astronomy, Shanghai Jiao Tong University, Shanghai 200240, China}

%\address[mymainaddress]{Department of Physics, Dezhou University, Dezhou, 253023, China}

\begin{abstract}
We represent a nonparametric method to reconstruct the equation of state for dark energy directly from observational Hubble parameter data. We use principal component analysis (PCA) to extract the signal from data with noise. Moreover, we modify Akaike information criterion (AIC) to guarantee the quality of reconstruction and avoid over-fitting simultaneously. The results show that our method is robust in reconstruction of dark energy equation of state. Although current observational Hubble parameter data alone can not give a strong constraint yet, our results indicate that future observations can significantly improve the quality of the reconstruction.
\end{abstract}

\begin{keyword}
\texttt{Dark Energy\sep Reconstruction\sep Nonparametric Model\sep Observational Hubble parameter Data\sep Principal Component Analysis}
%\MSC[2010] 00-01\sep  99-00
\end{keyword}

\end{frontmatter}

%\linenumbers

\section{Introduction}
\label{sec:intr}
Distance measurements of the type Ia supernovae (SNe Ia) indicate that the expansion of the universe is accelerating \cite{1999ApJ.517.565,1998AJ.116.1009,2001ApJ.560.49R}. This implies that some mechanism must exist to provide a repulsive effect. Many theories have been proposed to explain this repulsive effect. The most popular one is the dark energy paradigm. Dark energy is one kind of special matter which can provide the repulsive force that accelerates the expansion of the universe. However, its nature still remains unclear. To study the property of dark energy, Turner and White \cite{1997PRD.56.R4439} suggested to parameterize the dark energy by its equation of state, $w\equiv P/\rho$, where $P$ is pressure and $\rho$ is energy density. For different dark energy models, $w$ takes different values (e.g., -1 for vacuum energy, $-N/3$ for topological defects of dimensionality $N$), and $w$ may also evolve with time (e.g., models with a rolling scalar field). The analysis from the Planck + WP and BICEP2 data shows that the equation of state $w\neq-1$ and the bounce inflation scenario \cite{2004PRD.69.103520} is better than the standard cold dark matter model with a cosmological constant \cite{2013PRD.88.063501,2013PRD.88.063539,2014PRL.112.251301}. Although observation is consistent with $\Omega_{M} < 1$ and a cosmological constant $\Lambda > 0$ \cite{1990Nature.348.705}, the possibility of a time dependence of $w$ or a coupling with cold dark matter cannot be excluded \cite{2003PRD.67.063521}. Current observational data (SNe + Planck2015 + BAO) favor the cosmological constant model \cite{2016EPJC...76..588X} or XCDM and $\phi$CDM models \cite{2018arXiv180205571O}. In addition, on the theoretical level a constant $\Lambda$ runs into serious problems, since the present value of $\Lambda$ is $~10^{123}$ times smaller than the prediction from most particle physics models. If $\Lambda$ is not a constant, the dynamic properties of the dark energy may interest us \cite{2006IJMPD.15.2105,2001PRD.64.123527}. Since $w$ can well describe the dynamic properties of dark energy, it is vital to determine $w$ in the research of dark energy.

To determine the equation of state for dark energy, there are many different observational data sets that can be used. CMB anisotropy, supernovae (SNe) distance measurements and number counts all appear to be promising. One can fit directly to SNe magnitudes or their luminosity distances $d_{L}(z)$, or to more indirect quantities such as dark energy density, $\rho(z)$, the expansion history, $H(z)$ \cite{2012JCAP...02..048C}. But even accurate measurements of $d_{L}$ cannot constrain the small bumps and wiggles that are crucial to the reconstruction of $w$. Without some smoothing of the cosmological measurements, the reconstruction is unfeasible \cite{2001PRD.64.123527}. As the combination of number counts and supernova measurements could determine $H(z)$ directly and $H(z)$ could eliminate the dependence of the second derivative of $d_{L}$, using observational Hubble parameter data (OHD) to constrain $w$ seems to be a good choice \cite{liu2016PDU}.

Recently, there have been many works focusing on dark energy equation of state reconstruction. One can approach the reconstruction in a parametric or non-parametric way. Studying dark energy in a parameterized way, i.e., parameterizing the dark energy equation of state in terms of known observables \cite{2003PRD.67.063521}, e.g. Chevallier-Polarski-Linder parametrization \cite{2001IJMPD.10.213,2003PRL.90.091301} and divergency-free parametrizations \cite{2011PLB.699.233}, may induce biased results due to prior assumptions of function forms of the equation of state \cite{2013PRD.88.103528}. It may be more robust to reconstruct dark energy equation of state with non-parametric ways, since we do not know the nature of the dark energy so far. Reconstruction methods can also be divided into model dependent methods and model independent methods. Model dependent methods work within a particular model, while model independent methods have no such constraint. In fact, as a consequence of the strong degeneracy between cosmological models \cite{2014PhRvD..90d3531A}, any imposition would cause biased results \cite{2012PRD.86.123516}. Therefore, a reconstruction of $w(z)$ should be carried out ideally in a model independent manner \cite{2011MPLA.26.20}.

In this paper, we apply principal component analysis (PCA), which is a useful non-parametric model-independent tool, on the reconstruction of dark energy equation of state with observational Hubble parameter. However, in order to reconstruct $w(z)$ we need to calculate the derivatives in $H(z)$, which can, in general, increase the errors on the reconstructed equation of state \cite{2002PRD.65.103512,2013PRD.88.103528}. To overcome this deficiency, we try to fit $w(z)$ directly to OHD through the integral that relates $w(z)$ to $H(z)$. Another problem is that PCA can determine the principal components in $w(z)$, but how many components one should keep is still an open question. Too many components may cause overfitting and vise versa. So we propose to modify Akaike information criterion (AIC) in order to guarantee the quality of reconstruction and avoid over-fitting.

This paper is organized as follows. First, we describe the observational Hubble parameter data and explain the generation of simulated data in section~\ref{sec:met:simulate}. Next, we present the reconstruction process of dark energy equation of state in section~\ref{sec:met:reconstruction}. After that, we introduce the information criterion for components selection in section~\ref{sec:met:info}. Then we show results in section~\ref{sec:result}. Finally, we discuss the implications of our results and draw the conclusions in section~\ref{sec:conclusion}.

\section{Observational and  simulated data}
\label{sec:met:simulate}

\begin{figure}
\centering
\resizebox{1.0\textwidth}{!}{
\includegraphics{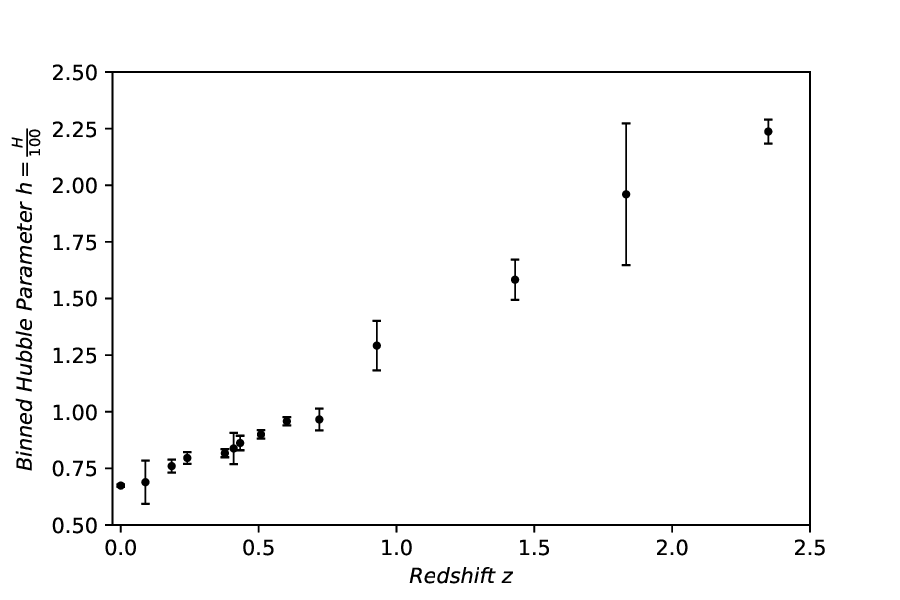}
}
\caption{ Binned observational Hubble parameter data. Black dots are binned observational Hubble parameter, which extend to $z=2.3$.  The bars show the deviation of Hubble parameter data, which is used to build an error model for the generation of simulated Hubble parameter data.}
\label{fig:databin}
\end{figure}

 Current observational hubble parameter data are obtained primarily from the method of cosmic chronometers \cite{2005PRD.71.123001,2010AJSS.188.280,2012JCAP.09.020,2014RAA....14.1221Z}. Other methods to extract $H(z)$ are by the observations of BAO peaks \cite{2009MNRAS.399.1663,2012MNRAS.425.405} and Ly-$\alpha$ forest of luminous red galaxies (LRGs) \cite{2013AA.552.A96}, which has extended the current OHD up to $z = 2.36$. Ref.\cite{cao2018a,cao2018b} provides the data set we use, which covers several independent measurements of $H(z)$. $H_0$ is set the value measured with Planck based on CMB \cite{2018arXiv180706209P}, i.e., $67.4 \pm 0.5 kms^{-1} Mpc^{-1}$. There are $44$ data points in the data set in total ($H_{0}$ included). As the redshift distribution of these data points is not uniform, we split the redshift region ($0<z<2.36$) into $13$ bins by the binning strategy in Ref.\cite{cao2018b}. The binned observational Hubble parameter data set is shown in figure~\ref{fig:databin}.

In order to validate the reconstruction process described in section~\ref{sec:met:reconstruction}, we use the simulated Hubble parameter data set, generated from some toy models (with known $w(z)$), to reconstruct the equation of state. Then we can evaluate the quality of reconstruction through comparing reconstructed equation of state with the known $w(z)$.

\begin{figure}[tbp]
\centering
\resizebox{1.0\textwidth}{!}{
\includegraphics{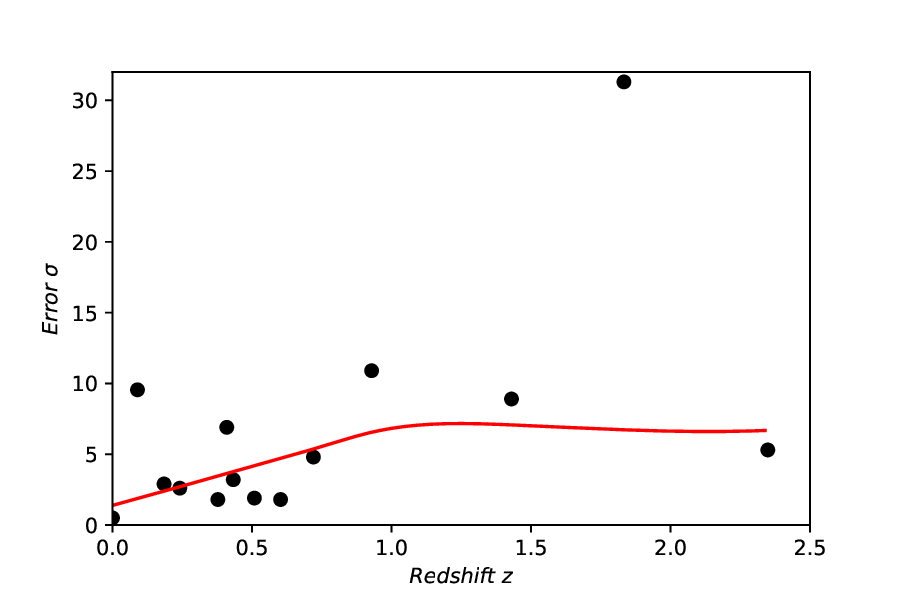}
}
\caption{ Error model obtained from OHD data. The red curve is obtained through the lowess regression fitting to the error of binned OHD data. The error of binned OHD data is plotted as the black dots.}
\label{fig:errormodel}
\end{figure}

To generate a simulated Hubble parameter data set, we need a fiducial model which can be characterized by an equation of state $w(z)$, from which we can get the theoretical Hubble parameter $H_{th}$ via Eq.~(\ref{eq:hth1}) with a particular $w(z)$. In addition to the underlying model, we also need an error model, which estimates the deviations for simulated Hubble parameter data from the theoretical values. There have been some studies on how to obtain an error model from observational Hubble parameter data. For instance, Yu et al.\cite{2013PRD.88.103528} suggested that the error of observational Hubble parameter data follow Nakagami $m$ distribution; Ma et al. \cite{2011ApJ.730.74} used the center line of upper heuristic bounds and lower heuristic bounds of the observational Hubble parameter data as the error model for simulated Hubble parameter data. Generally, the uncertainty of the observational data is strongly dependent on redshift. In this study we generate the error model for simulated Hubble parameter data by fitting the error of the observational Hubble parameter data with a lowess regression, as shown in figure~\ref{fig:errormodel}.

Finally, with the fiducial model and the error model discussed above, we can generate the simulated Hubble parameter data via

\begin {equation}
\label{eq:hsim}
H_{sim}(z) = H_{th}(z) + G(0,\sigma(z)),
\end {equation}
where $G(0,\sigma)$ is the Gaussian distribution with zero mean and the standard deviation of the distribution $\sigma(z)$ is the lowess regression curve.

For each fiducial model we generate 14 simulated $H(z)$ data points, the same number as binned OHD, with $z$ being uniformly distributed in $[0,2.3]$ by step 0.177.

\section{Reconstructing process}
\label{sec:met:reconstruction}

We reconstruct the equation of state for dark energy with the assumption that the universe is homogeneous, isotropic and governed by Einstein's theory of gravitation. It is well known that the metric for a space-time with homogeneous and isotropic spatial sections is the maximally-symmetric Friedmann-Robertson-Walker (FRW) metric. We also treat the whole universe as ideal fluid. Then from Friedmann equation we have

\begin{equation}
\label{eq:hth0}
\frac{H^{2}(z)}{H^{2}_{0}}= \Omega_{m}(1+z)^{3}+\Omega_{r}(1+z)^{4}+\Omega_{x}e^{3\int^{z}_{0}\frac{1+w(z\prime)}{1+z\prime}dz\prime}+\Omega_{k}(1+z)^{2},
\end{equation}
where $\Omega_{x}$ is dark energy density component, $\Omega_{m}$ is matter density component, $\Omega_{k}$ is curvature component, and $\Omega_{r}$ is radiation density component respectively. According to the latest result from Plank, we adopt $H_{0}=(67.4\pm0.5)km s^{-1}Mpc^{-1},\Omega_{x}=0.685,\Omega_{m}=0.315,\Omega_{k}=0,\Omega_{r}=0$ \cite{2018arXiv180706209P}. Then Eq.~\ref{eq:hth0} can be reduced to

\begin {equation}
\label{eq:hth1}
\frac{H^{2}(z)}{H^{2}_{0}}=\Omega_{m}(1+z)^{3} + \Omega_{x}e^{3\int^{z}_{0} 1+w(z\prime)d\ln{(1+z\prime)}},
\end {equation}
from which we can calculate the theoretical Hubble parameter $H_{th}(z)$ when the expression of $w(z)$ is known. We can fit observational $H_{ob}(z)$ data to constrain the analytical form of $w(z)$ through least-square method,

\begin {equation}
\label{eq:chisquare}
\chi^{2}=\sum_{i=1}^{N} {\frac{[H_{ob}^{i}(z)-H_{th}^{i}(z)]^{2}}{\sigma_{i}^{2}}},
\end {equation}
where $N$ is the number of data points, $H_{ob}$ is observational or simulated Hubble parameter data, $H_{th}$ is theoretical Hubble parameter, and $\sigma$ is the error of $H_{ob}$ respectively. However, it is well known that applying Least-square minimization can not distinguish between noise and signal, as noise is included in the result of optimization. In order to minimize the effect of noise, we use PCA to determine the principal components of $w(z)$.

The process of applying PCA on reconstruction of equation of state is described as follows:

Step 1. We choose a set of basis functions $f_{i}(z)=z^{i-1},i=1,2,...,K$ and equation of state $w(z)$ can be expressed as

\begin {equation}
\label{eq:wparameterize}
w(z)=\sum_{i=1}^{K}\alpha_{i}f_{i}(z),
\end {equation}
where $\alpha_{i},i=1,2,...,K$ are the coefficients of the corresponding basis functions.

Step 2. We use Eq.(~\ref{eq:hth1}) and (\ref{eq:chisquare}) and non-linear least-square optimization to derive the coefficients $\alpha_{i}$.
%so we can obtain the distribution of $\alpha_i$ with enough runs of step 2, each run with $H(z)$ data simulated from the error data.

Step 3. We simulate $H(z)$ enough times based on the error data (e.g. see the error bars in Fig.\ref{fig:databin}), each simulation produces a different set of  $H(z)$ data. For each set of $H(z)$ data we apply Step 2 and get one instance of $\alpha_{i}$. Thus we can get enough instances of $\alpha_{i}$, from which we calculate the covariance matrix of $\alpha_i$, denoted by C.

Step 4. We calculate the eigenvalues ${\lambda_{i}}$ and eigenvectors by diagonalizing the covariance matrix C,

\begin {equation}
\label{eq:covariance}
C = E \Lambda E^{T},
\end {equation}
where $\Lambda$ is the diagonal matrix. The diagonal entries of $\Lambda$ are eigenvalues, and the columns of matrix $E$ are corresponding eigenvectors.

Step 5. We sort the components by $\lambda_{i}$ in ascending order and choose first $M$ components. That is, the $K \times K$ matrix $E$ is reduced to dimension $K \times M$. Then we define a new basis $U=F E$, where $F=(f_{1},f_{2},...,f_{K})$. This new basis $U=(u_{1},u_{2},...,u_{M})$ are linear combinations of the original basis functions $f_{i}$, and they are orthogonal. The $u_{i}$s corresponding to small eigenvalues are the relatively stable components, which vary little for different observations. They are hence the components we need.

Step 6. Based on Eq.(\ref{eq:hth1}) and (\ref{eq:chisquare}), we use the new basis functions $U=(u_{1},u_{2},...,u_{M})$ and non-linear Least-square Optimization to fit $H(z)$ data again and obtain the new coefficients $\beta_{i}$. Finally we get equation of state
\begin {equation}
\label{eq:wconstructed}
w(z) = \sum^{M}_{i=1}{\beta_{i}u_{i}(z)},
\end {equation}
where $M$ is the number of used components. The covariance matrix of $\beta_{i}$ is just the diagonal matrix $\Lambda$ derived from Eq.(\ref{eq:covariance}). Besides, a criterion is needed to decide how many components (i.e. the value of $M$) should be used, which will be discussed in section~\ref{sec:met:info}.

\section{Criterion for components selection}
\label{sec:met:info}

We have discussed the reconstruction process in section~\ref{sec:met:reconstruction}. Then we need a criterion to determine how many components we should keep in the reconstruction. This is still an open question in principal components analysis. Many criteria have been proposed to solve this problem since PCA was invented in 1901, e.g. Akaike information criterion (AIC), Bayesian information criterion (BIC) \cite{2007MNRAS.377.L74}, and the combinations of them. Yu et al. used a criterion called Goodness of Fit \cite{2013PRD.88.103528}.  Although these criteria works well in general, there are defects in these criteria: introducing either additional parameter or additional assumptions, such as $s$ in the combination of AIC and BIC \cite{2010PRL.104.211301}. Furthermore, evaluating these parameters arbitrarily may lead to quite different reconstructing results in some particular problems.

In this paper, we introduce a new criterion for principal components selection. It is a modified AICc (the small-sample-size corrected version of AIC) criterion. The expression of AICc is \cite{1989Biometrika,1997statistics33}

\begin{equation}
\label{eq:aicc}
AICc = \chi_{min}^{2} + (\frac{2N}{N-M-1})M,
\end{equation}
where $M$ is the number of parameters, i.e. the number of components we keep, and $N$ is the size of data set. $\chi_{min}^{2}$, i.e. Eq.(\ref{eq:chisquare}), represents the deviation from observational data or simulated data. It is expected that reducing $M$ will be accompanied by a reduction in the error, but an increased chance of getting $w(z)$ wrong \cite{2010PRL.104.211301}, which means $w(z)$ is reconstructed less accurately (so the bias increases), but the error bars are smaller (so the variance decreases)\cite{2003PRL.90.031301}. It turns out that the challenge for this issue is to achieve a balance between the bias and variance. Therefore we add an empirical parameter $s$ into the latter term of AICc to determine the value of $M$. After that, the modified AICc criterion is

\begin{equation}
\label{eq:aiccm}
AICcm = \chi_{min}^{2} + s(\frac{2N}{N-M-1})M,
\end{equation}
where $s$ is the tuning factor we add to AICc to include the penalty for more parameters and to avoid overfitting.

The introduction of parameter $s$ endows the criterion with additional flexibility, which makes it adaptable to specific problems. The value of $s$ is determined empirically and consistent with the fact that results of best fit have the minimum AICcm values. Theoretically, for the best $M$ that corresponds to minimum AICcm, the derivative of AICcm with respect to $M$ should be zero. However, this point is seldom exactly satisfied since $M$ is confined to integers. Nevertheless, we can determine the value of $s$ approximately with the help of AICcm derivative with respect to $M$, which is shown as follows.

To stress the dependency on $M$ we rewrite Eq.(\ref{eq:aiccm}) as

\begin{equation}
\label{eq:aiccm2}
  AICcm(M)=\chi_{min}^{2}(M) + sA(M)
\end{equation}
where $A(M)=\frac{2NM}{N-M-1}$. Then for one fixed $M$ we can get
\begin{equation}
\label{eq:dplusaiccm}
\begin{split}
  \Delta_{+}AICcm(M) &=AICcm(M+1)-AICcm(M) \\
  &=\Delta_{+}\chi_{min}^{2}(M)+s\Delta_{+}A(M) \\
  &=\chi_{min}^{2}(M+1)-\chi_{min}^{2}(M)\\
  &  +s[A(M+1)-A(M)]
  \end{split}
\end{equation}

 $\Delta_{+}AICcm(M)$ can be taken as the proxy of right derivative of $AICcm(M)$ with respect to $M$. Since $AICcm(M)$ get minimum at the fixed $M$ we let $\Delta_{+}AICcm(M)=0$ and get

 \begin{equation}
\label{eq:splus}
  s_{+}=-\frac{\Delta_{+}\chi_{min}^{2}(M)}{\Delta_{+}A(M)}=-\frac{\chi_{min}^{2}(M+1)-\chi_{min}^{2}(M)}{A(M+1)-A(M)}
\end{equation}
Correspondingly,

 \begin{equation}
\label{eq:sminus}
  s_{-}=-\frac{\Delta_{-}\chi_{min}^{2}(M)}{\Delta_{-}A(M)}=-\frac{\chi_{min}^{2}(M)-\chi_{min}^{2}(M-1)}{A(M)-A(M-1)}
\end{equation}
Then the value of $s$ is calculated by

 \begin{equation}
\label{eq:svalue}
  s=(s_{+}+s_{-})/2
\end{equation}

\begin{table*}
\centering
\caption{\label{tab:srange} The Values of $AICcm$ , $M$ and $s$  for Different Models}
\begin{tabular}{cccc}
\hline \hline
  $Model$& $AICcm$ & $M$ &$ s $\\
\hline
$\Lambda CDM$ & $2.6547$ & $4$ & $0.2028$ \\
$model \ 1$ & $4.6730$ & $4$ & $0.3604$ \\
$model \ 2$ & $2.6710$ & $4$ & $0.2108$ \\
$OHD    $   & $37.6465$ & $3$ & $3.6208$ \\
\hline \hline
\end{tabular}
\end{table*}

%Before reconstructing the equation of state for the fiducial models, we need obtain the optimum value of $s$ with the method described in section~\ref{sec:met:info}.
Hence, once the best number of components $M$ is fixed, we can calculate s through Eq.~(\ref{eq:splus}-\ref{eq:svalue}). Now we can see that the value of s and the best $M$ depends on each other: In order to calculate $s$, the best $M$ need to be given, while the determination of the best $M$ takes $s$ as the prerequisite. To address this dilemma, we designed a iterative procedure that alteratively computes $s$ and $M$. We initialize this procedure by set $s=1$, which corresponds to AICc criterion shown in Eq.~(\ref{eq:aicc}). In practice, this procedure can converge within 10 iterations generally. The values of $AICcm$ , $s$ and the best $M$ for different $w(z)$ models and OHD are listed in table 1.
%The $AICcm$ values for the above three models are shown in Tab.~\ref{tab:aicc} and the ranges of $s$ are shown in Tab.~\ref{tab:srange}. The optimal value of $s$ we get is $0.02753$. Having $s$ fixed, we can apply the $AICcm$ criteria on components selection. In other words, we should pick up proper number of components to make the $AICcm$ achieves the minimum.

\section{Results}
\label{sec:result}

\begin{figure}[tbp]
\centering
\includegraphics[width = 1.0\textwidth]{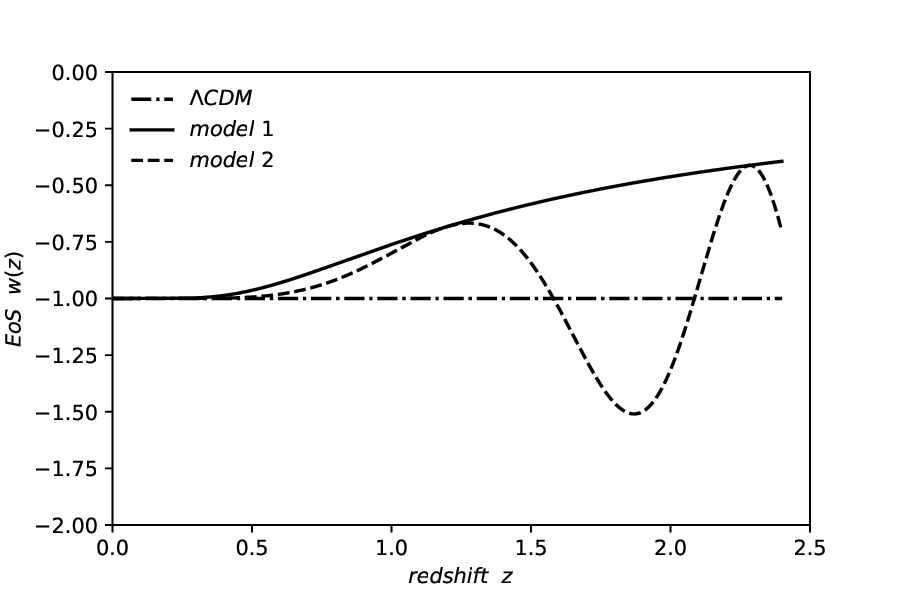}
\caption{Equation of state for three models. Solid line shows the equation of state of $\Lambda CDM$ model, dash line shows the equation of state of model-1, and dash-dot line shows the equation of state of model-2. For model-1, $w(z)$ is $-1$ at $z=0$ and approches to $0$ as redshift increases. For model-2, $w(z)$ oscillates around $-1$ with an amplitude increases from $0$ to $1$.}
\label{fig:3eos}
\end{figure}

Following the reconstruction process described in section~\ref{sec:met:reconstruction}, we first reconstruct the equation of state from simulated Hubble parameter data generated from the fiducial model (with pre-set equation of state $w(z)$). There are three $w(z)$ we reconstruct: a. $\Lambda CDM$ model, $w(z) = -1$; b. model-1, $w(z) = -\tanh(\frac{1}{z})$; c. model-2, $w(z) = -1 + [1-tan(\frac{1}{z})]\sin{z^{2.5}}$. Their equations of state are shown in figure~\ref{fig:3eos}.

\begin{figure}[tbp]
\centering
\includegraphics[width = 1.1\textwidth]{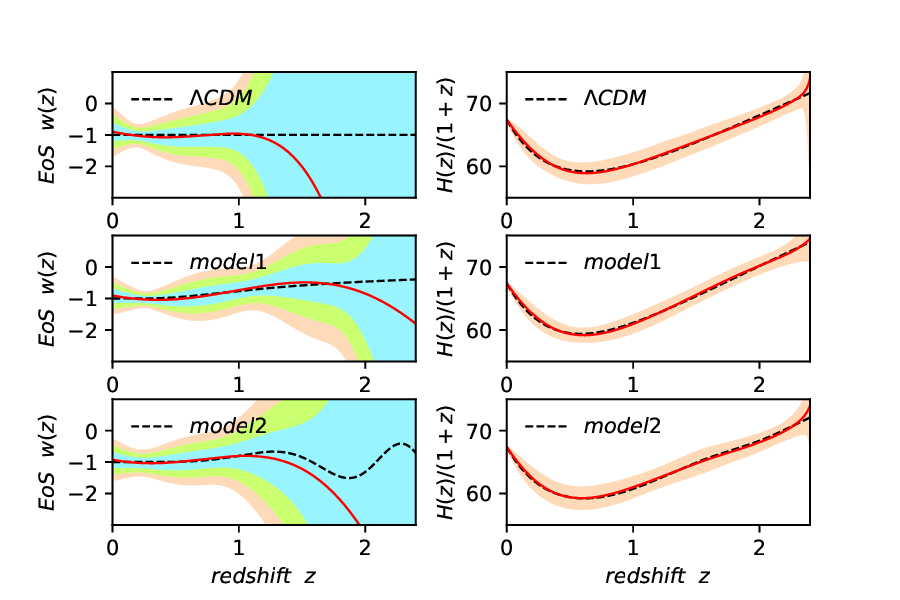}
\caption{ (color online). Left column: Reconstructed $w(z)$ for different models. The light blue shade, green shade and orange shade shows 1$\sigma$, 2$\sigma$ and 3$\sigma$ error respectively. Black dash line is pre-set $w(z)$, and red solid line shows the reconstructed $w(z)$. Right column: Hubble parameter $H(z)/(1+z)$. Red solid line is the Hubble parameter calculated from the reconstructed $w(z)$. The orange shade shows its 1$\sigma$ error, and black dashed line shows the Hubble parameter calculated from pre-set $w(z)$.}
\label{fig:3model}
\end{figure}

The reconstruction results are shown in figure~\ref{fig:3model}. It is evident that the reconstruction is reliable. Pre-set $w(z)$s are well in $1 \sigma$ regions of reconstructed $w(z)$s. Furthermore, the reconstructed $w(z)$ well recovers the trends of pre-set $w(z)$s up to 1.5 or so. In terms of variance, we can see the variances of reconstructed $w(z)$s are under control at $0<z<1.5$. But at $z>1.5$, the variances dramatically increase. The behavior of reconstructed $w(z)$ diffs to some extent among different models. The right panel in figure~\ref{fig:3model} shows that the Hubble parameter (for convenience, we draw $H(z)/(1+z)$ instead of $H(z)$) calculated from reconstructed $w(z)$ is consistent with the Hubble parameter obtained from pre-set $w(z)$s. And the orange region is narrow, which implies that the variance of reconstructed Hubble parameter is very small.

\begin{figure}[tbp]
\centering
\resizebox{1.1\textwidth}{!}{
\includegraphics[width = 1.1\textwidth]{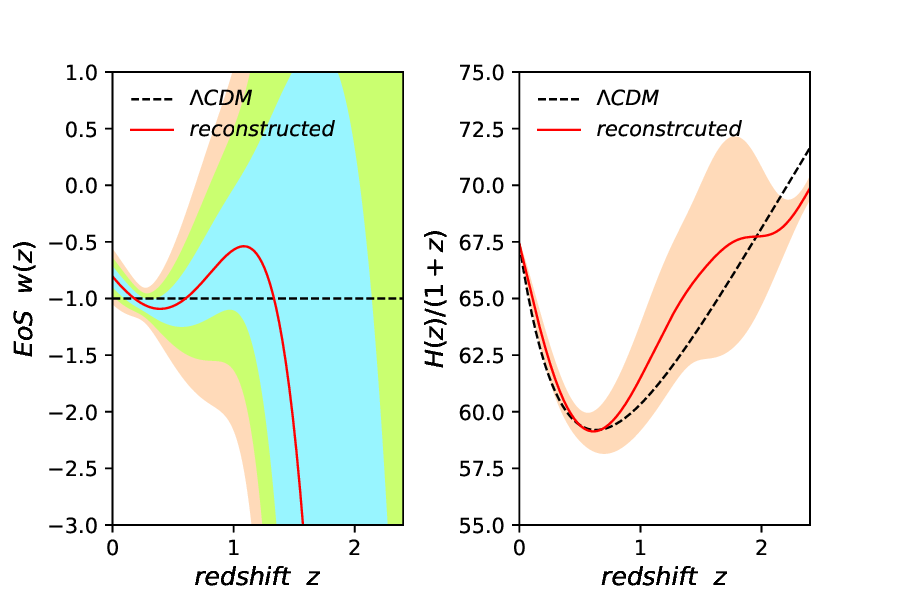}
}
\caption{ (color online). Left: The reconstructed $w(z)$ from OHD data (with no underlying model). The light blue shade, green shade and orange shade shows 1$\sigma$, 2$\sigma$ and 3$\sigma$ error respectively. Right: Red solid line is the $H(z)$ calculated from the OHD-reconstructed $w(z)$. Black dashed line is the $H(z)$ calculated from $w=-1$.}
\label{fig:OHD}
\end{figure}

Then we reconstruct $w(z)$ from the observational Hubble parameter data without underlying model, which is shown in Figure~\ref{fig:OHD}. We can see that the reconstructed $w(z)$ oscillates around $w=-1$ at $0<z<1.5$, but at $z>1.5$ it deviates from $w=-1$ rapidly, and the variance of it is large. Compared to the reconstructed $w(z)$ from the simulated Hubble parameter data (Fig.~\ref{fig:3model}), the OHD-reconstructed $w(z)$ oscillates somewhat more wildly. This may be due to the fact that, compared to the observational data, the simulated redshifts $z$ are uniformly sampled and the theory values of simulated Hubble parameters, computed from the underlying $w(z)$ model directly (through Eq.(\ref{eq:hth1})), monotonically increase with $z$. Nevertheless, combining the observational and simulated results we can conservatively conclude that the reconstruction is robust at $0<z<1.0$.
%The right panel in figure~\ref{fig:OHD} shows the reconstructed Hubble parameter. It is clear that although the Hubble parameter calculated from the reconstructed $w(z)$ of simulated Hubble parameter data (with model-1) is consistent with the theoretical Hubble parameter of model-1, its variance is very large. Figure~\ref{fig:OHD} also shows the Hubble parameter calculated from $w_{reco}$, which is out of 1$\sigma$ region of Hubble parameter calculated from reconstructed $w(z)$ of the simulated Hubble parameter data (with model-1) and deviates largely from real observational Hubble parameter data.

\section{Conclusions and discussion}
\label{sec:conclusion}

We apply principal component analysis to reconstruct the dark energy equation of state. We represent $w(z)$ as linear combination of a set of orthogonal basis functions to avoid information loss. In addition, we modify the corrected Akaike information criterion (AICc) to balance between bias and deviation in the reconstruction. Then we reconstruct the equation of state for three models from simulated Hubble parameter data to validate our reconstruction process. The results show that we can constrain the equation of state quite well at $0<z<1.5$ or so, and the Hubble parameter calculated from reconstructed $w(z)$ is consistent with the Hubble parameter calculated from pre-set $w(z)$ very well as the 1$\sigma$ region is very narrow. However, differences of the Hubble parameter derived from different models are very small, although the models are quite different. This is also the reason why reconstructing dark energy equation of state from Hubble parameter data is difficult. Nevertheless, from this study we show that future observations with larger quantity and higher quality of observational Hubtle parameter data will help to constrain dark energy equation of state to redshift $z \simeq 2.0$.

Higher-redshift, more-complete-sky-coverage LRG survey and spectroscopic observations of those identified LRGs will remarkably enhance the quality of observational Hubble data. Considering a future LRG survey with more than half sky coverage, such as 2SLAQ\footnote{http://www.2slaq.info/}, it may give several millions of LRGs and effectively enlarges the number of OHD data points or lowers their errors. We confirm that the ongoing CMB observation program, the Atacama Cosmology Telescope (ACT)~\cite{actweb}, and the forthcoming surveys such as Euclid \cite{Laureijs2011} and WFIRST \cite{Spergel2013} will greatly improve the present-day statistics and produce a promising dataset of larger quantity and better quality, which will result in greatly improved reconstruction of dark energy \cite{2013PRD.88.103528}. Furthermore, a small sample collected at much higher redshift, which may be implemented in near future, will reduce the errors more efficiently than collecting more events in the original redshift interval \cite{2001PLB.500.8}.

%In the case of reconstructing dark energy equation of state from real observational Hubble parameter data, we can constrain $w(z)$ at $0<z<1.0$ with an underlying model. But the reconstructed $w(z)$ deviates largely from $w(z)$ of underlying model at $z>1.0$, as the quality of data is very poor. If there is no underlying model, the reconstructed $w(z)$ oscillates even more wildly, and we can not know whether it is close to the true $w(z)$.

%Since we do not understand the nature of dark energy, it is possible that dark energy can lead to other observable effects such as a new long range force \cite{1998PRL.81.3067}. More accurate observations are needed to reveal the secrets of dark energy.

\section*{Acknowledgments}
We thank Ying-Jie Peng for helpful comments and discussions. This work was supported by the National Science Foundation of China (Grants No. 11573006, 11528306), the National Key R\&D Program of China (2017YFA0402600), the Natural Science Foundation of Shandong Province, China(Grant NO.ZR2019MA059), the Fundamental Research Funds for the Central Universities and the Special Program for Applied Research on Super Computation of the NSFC-Guangdong Joint Fund (the second phase).
\bibliographystyle{elsarticle-num}
\bibliography{references}

\end{document}